# Is Your Load Generator Launching Web Requests in Bunches?

James F Brady
Capacity Planner for the State of Nevada
jfbrady@admin.nv.gov

*One problem with load test quality, almost always overlooked, is the potential for the load generator's user thread pool to sync up and dispatch queries in bunches rather than independently from each other like real users initiate their requests. A spiky launch pattern misrepresents workload flow as well as yields erroneous application response time statistics. This paper describes what a real user request timing pattern looks like, illustrates how to identify it in the load generation environment, and exercises a free downloadable tool which measures how well the load generator is mimicking the timing pattern of real web user requests.*

**1.0 Introduction**

Most web application load testing professionals assume their load generator's virtual user thread pool precisely mimics the request timing that a comparable set of real users produce. A traffic generator, however, is one computer initiating web requests with a large fixed set of user threads operating in closed loops while real users each have their own computing device and make queries independently from each other as a dynamically changing subset of a larger population. The simulator's heavier workload, the think time method it uses, and the feedback produced by its fixed closed loops may cause the user thread pool to sync up and offer unrealistic surges of traffic to the system under test (**SUT**). Few practitioners think about user thread synchronization, and those that do find the problem difficult to quantify when it occurs. This paper describes the request pattern produced by real users and provides a measurement methodology for evaluating request pattern quality. The approach taken is illustrated with a free downloadable tool this author developed, the **web-generator-toolkit**.

The first step in this process, contained in **Section 2**, is to identify the request timing characteristics of real web users through a graphical illustration and simulation. With that foundation, **Section 3** describes how a single load generator attempting to mimic the request pattern of many real users can have its user process threads sync up and launch queries in bunches. The **Section 2** and **Section 3** information is applied in **Section 4** to an example load test using the web request timing evaluation features of the **web-generator-toolkit**. Data representing real user request timing are analyzed along with data that does not possess those properties. **Section 5** wraps up the discussion with some conclusions and summary remarks. As a convenience, links are setup in this document for **references**, **definitions**, and **section** / **figure** locations. To return from internal links use the Alt + left arrow keys.

**2.0 Real User Vs Virtual User Request Timing**

The physical difference between real users and load tool based virtual users querying a web application is illustrated in **Figure 1**. The three real users on the left are shown as separate entities with their own computing device and communications connection to the server, while the three virtual users on the right are sharing one computing device and an associated communications connection. Each real user is encapsulated within its own box of activity independently querying the application without coordinating web page choice or request timing with each other. In contrast, the three virtual users share the same box of user activity so their web page choice may follow a pattern and the timing of their queries could become collectively synchronized. The box around the middle real user is dotted to indicate that he has just transitioned into the active user pool and is making his first web request. The other two real users are maintaining their current session but could just as easily be transitioning out of the active user pool. The virtual user thread pool on the right remains constant in size with each thread fixed in position and operating in a closed loop.

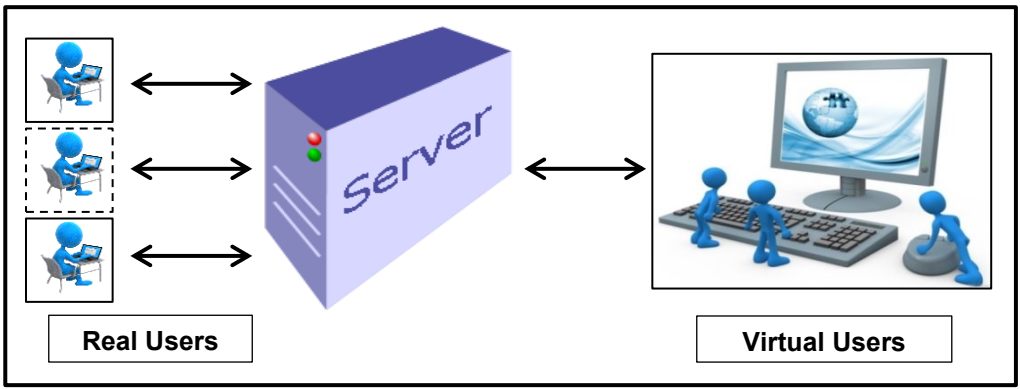

**Figure 1: Real Users Vs Virtual Users**

What does the query timing of the collective group of real users look like and can the load generator's request timing be captured to evaluate how well it is matching that timing? The place to begin is the timing of the queries initiated by the members of the real user pool actively engaged with the application.

**2.1 Request Timing Illustration**

All three real web users in **Figure 2** are engaged with the application but in a completely uncoordinated manner relative to each other. The top user is receiving a home page response, while the middle user is initiating a login, and the bottom user is thinking about what to request next. Note, the user's access device timestamp is the one of interest, not a timestamp recorded on the target server when receiving the query.

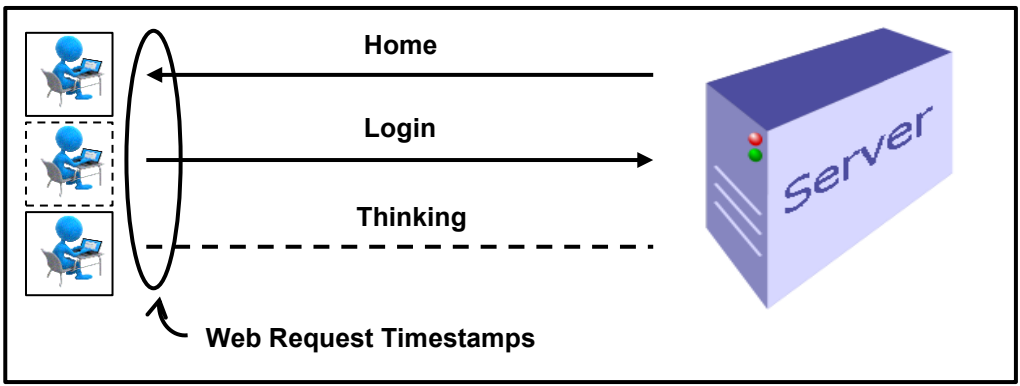

**Figure 2: Real User Web Requests**

So, what does the set of query start times look like on a time line for the combined pool of active users? Is the timing of queries an evenly spaced "assembly line" of events or some other pattern? It can be argued that the time line being analyzed has "no pattern at all" since individual queries are uncorrelated in time and therefore, "independent" from each other. **Figure 3** is a fragment of such a time line where the red diamonds represent query start times over seven ten second time intervals. Queries five and six were initiated by a pair of users at the same exact time, a likely occurrence since the users are operating independently from each other.

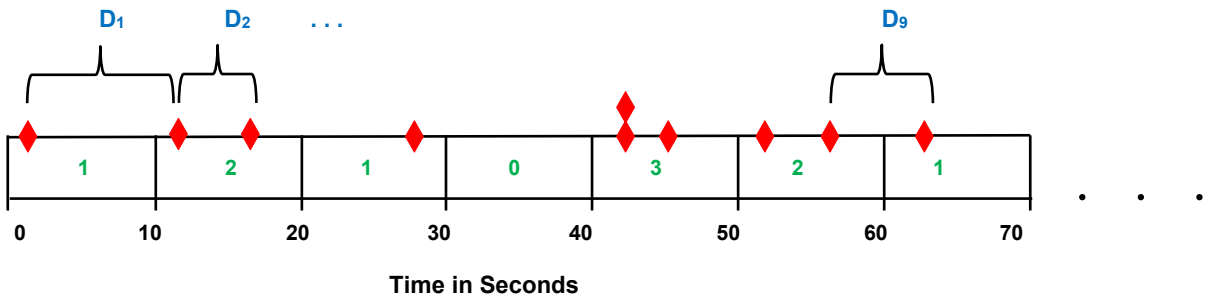

**Figure 3: Real User Request Time Line Fragment**

Query independence doesn't mean that request timing cannot be statistically profiled. In fact, under rather broad conditions this time series can be characterized by two probability distributions, one depicts the number of queries per constant length time interval and the other represents the time interval between queries. The green histogram in **Figure 4a** illustrates count per interval using the seven ten second periods in the **Figure 3** time-line fragment. For example, there is one ten second interval with no red diamonds and three of them contain one red diamond.

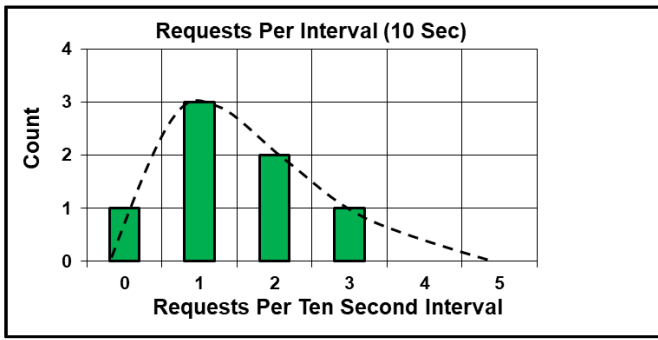 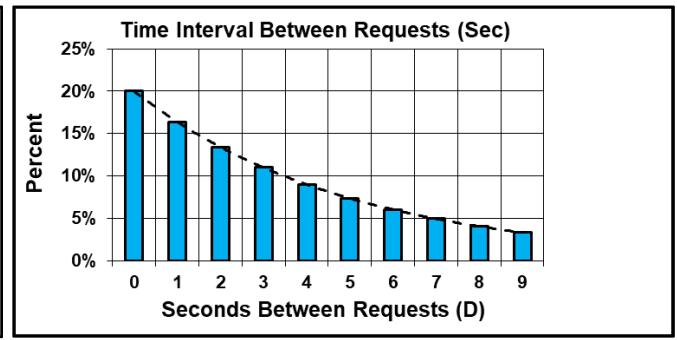

Figure 4a: Requests Per Interval (10 Sec)    Figure 4b: Time Interval Between Requests (Sec)

This fragment is too small to show a meaningful histogram of the time interval between queries so the blue histogram in **Figure 4b** is a full time-line illustration of those intervals where queries are initiated "**D**" seconds apart. Queries five and six are one pair of red diamonds in **Figure 3** that contribute to the **Figure 4b** "0" seconds between requests column because they are initiated at the same time.

**2.2 Request Timing Simulation**

With these basic ideas in mind, **Figure 5** and **Figure 6** represent a full measurement period simulation of these independent request ideas. **Figure 5** is a one-thousand second timeline laid out in five two-hundred second intervals with two hundred queries occurring during the one-thousand seconds. These simulated queries are produced using a random number generator which has an equal chance of returning any value between zero and one-thousand. The two-hundred values selected are plotted as red diamonds sectioned off in ten second intervals, like they are in **Figure 3**. For further details regarding this simulation see [**BRAD09**].

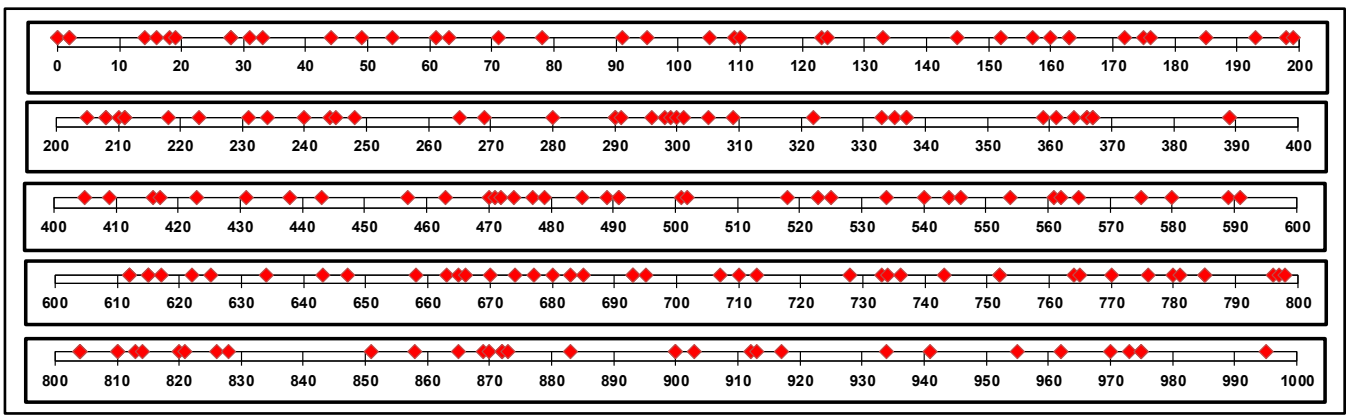

**Figure 5: Real User Request Time Line Over a 1000 Second Measurement Period**

The two graphs in **Figure 6** below are comparable to **Figure 4a** and **Figure 4b** with the vertical axis of **Figure 6a** expressed as a percent instead of the **Figure 4a** count. The time series of red diamond events in **Figure 5** is statistically summarized as the **Figure 6a** requests per ten second interval histogram and the **Figure 6b** seconds between requests histogram.

An inspection of **Figure 5** reveals how this mapping works. There are thirteen of the one-hundred ten second intervals with no red diamonds which is reflected in the **Figure 6a** "0" green column (13 / 100 = 13%). Also, there are only one-hundred and seventy-four red diamonds visible of the two hundred present because twenty-six of them occur at the same time and are, therefore, on top of each other. The **Figure 6b** blue "0" column graphically illustrates this information (26 / 200 = 13%).

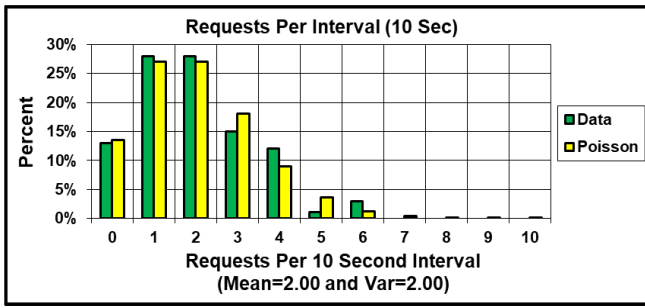
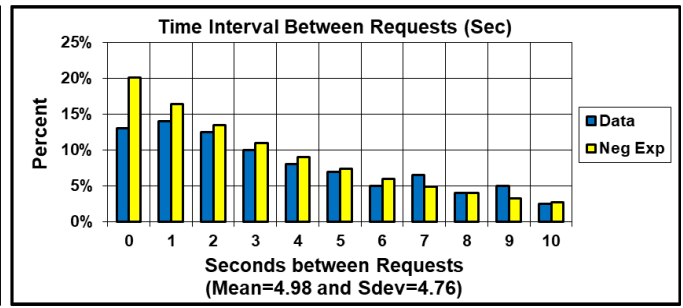

Figure 6a: Requests Per Interval (10 Sec)     Figure 6b: Time Interval Between Requests (Sec)

When events are drawn independently from each other, as they are here, the result is referred to in the literature as a **Poisson Process** characterized by the number of events per constant length interval being **Poisson** distributed and the time interval between those events possessing **Negative-Exponential** distribution properties. The yellow column charts in **Figure 6a** and **Figure 6b** are these theoretical equivalents of the green and blue histograms. The green column chart seems to match its yellow theoretical counterpart better than the blue chart fits to its, but they are both reasonably close.

Note that the **mean** of the requests per interval in **Figure 6a** is 2.00 and the variance is also 2.00. Likewise, the **mean** of the time interval between requests in **Figure 6b** is 4.98 and the **standard deviation**, or square root of the variance, is 4.76. It turns out that the **Poisson** distribution's **mean** equals its variance and the **Negative-Exponential** distribution's **mean** equals its **standard deviation**. The data in **Figure 5** comes very close to producing these theoretical numbers (**mean** request per interval: 200 requests / 100 intervals = 2 requests / interval and **mean** time interval between requests: 1000 seconds / 200 requests = 5 seconds / request).

## 2.3 Request Timing Evaluation

How can these statistical properties of independent requests be exploited to determine request pattern quality? The statistical properties of either the **Poisson** or the **Negative-Exponential** distribution can be used to make that determination. For **Poisson** distributed requests per interval, the analysis requires the data to be grouped into arbitrary fixed length intervals like the ten seconds used in **Figure 3** and **Figure 5** with **mean** and variance calculated on that grouped data basis. A simpler and more robust approach is to determine if the time intervals between requests are **Negative-Exponentially** distributed by calculating the **mean** and **standard deviation** of sorted timestamp differences to see how close the two numbers are to each other. Better yet, calculate the Coefficient of Variation (**CoV**) which is the **standard deviation** / **mean** ratio of these timestamp differences and see how near it is to one. If the **CoV** is close to 1.0 a real user request pattern is being generated, but if not, requests are being launched in bunches when the **CoV** is significantly greater than 1.0 and are too evenly spaced if the **CoV** is markedly less than 1.0. This leads to the following guiding principle when deciding if the load generator is producing real web user request timing.

> **Real Web User Request Timing**
>
> To determine if the load generator is creating web requests independently from each other like real users, sort the launch times in ascending order, calculate their differences, and compute the Coefficient of Variation (**CoV**) of those differences. If the **CoV** is approximately equal to one, **CoV** ~ 1.0, real user request timing is being produced. If the **CoV** > 1.0 requests are being launched in bunches and if the **CoV** < 1.0 they are too evenly spaced.

The **CoV** for time intervals between requests is the key to determining how independently the load generator is launching requests but is seldom computed because most load testing professionals are unaware that this important traffic quality measure exists. The **web-generator-toolkit** performs this calculation with a **JMeter** "Aggregate Report" csv file that contains rows of event data which include request timestamps. The timestamps are sorted in ascending order and **CoV** computed.

The next section steps through three circumstances when a load generator comes up short as a substitute for real users where the **CoV** can help the analyst identify the problem when it occurs.

### 3.0 Load Generator Launch Time Distortions

Armed with the above guiding principle and the tools for its implementation, a look is taken at some of the ways launch time distortions can occur focusing on three of the most common situations where time interval between requests **CoV** > 1.0. The three listed in the introduction expanded upon here are;
1. heavy workload,
2. think time method used, and
3. closed loop feedback.

The more "assembly line" oriented pattern of launched events with a **CoV** < 1.0 is illustrated in **Figure 15** and **Figure 16** of the **Section 4** example load test.

### 3.1 Heavy Workload

A load generating computer is an inherently synchronous device attempting to produce asynchronous requests. Computers use queues and scheduling to manage workload flow and both tend to synchronize the processing of user requests. For example, the think times (**Z**) drawn using the load tool's available timer options may represent a truly independent spacing of times between requests but, as **Figure 7** illustrates, the load generator implements that request by suspending the user thread with a sleep system call. When time is up the user thread does not execute immediately but is put on the run queue at its assigned priority and waits (W) in that queue before executing (E). Once the thread executes, it is suspended while waiting for the response (**R**) from the system under test (**SUT**). When the response arrives, the thread is again put on the run queue at its assigned priority and waits (W) before it executes (E). During execution it writes out its response information, sets up for the next request, and goes back to sleep (**Z**).

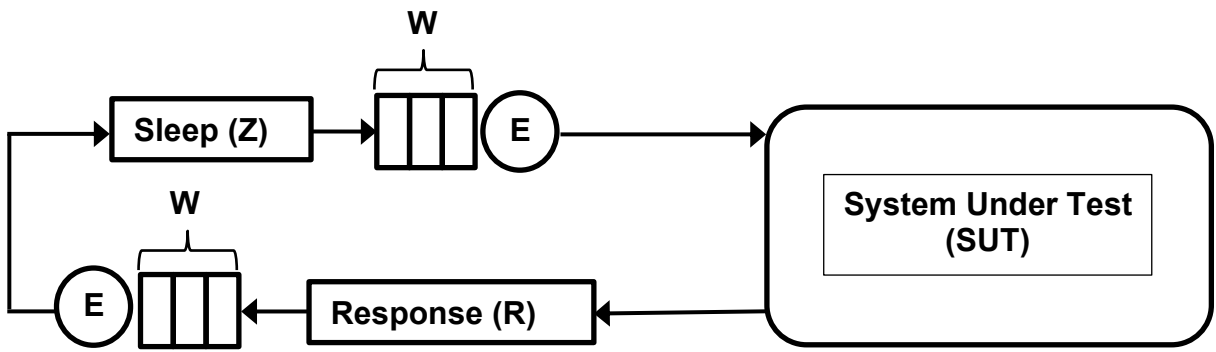

Figure 7: User Thread Launch Sequence

Under heavy load there can be significant launch time bunching caused by the large volume of events in the (W) and (E) states. In general, the sharing of processing execution resources increases contention for CPU cycles, adds to operating system run list size, and magnifies process thread scheduling complexity.

This problem can be mitigated by lightly loading a traffic generator that is configured with a lot of processor cores and real memory. If multiple lightly loaded traffic generators are needed, the output files they produce can be merged to evaluate request timing quality when their clocks are synchronized using the Network Time Protocol (**NTP**).

### 3.2 Think Time Method Used

The think time method used can also be a source of launch time distortion. The choice that usually causes this problem is fixed think times because, under those conditions, user threads are more likely to queue up and cluster behind the long response time events. Drawing think times from a probability distribution reduces the severity of this "lining up behind the long latency tasks in the test suite" situation.

In fact, the desired independent request timing can be achieved using any probability distribution if the number of user threads is sufficiently large. This "Principle of Superposition", [**ALBI82**] and [**KARL75**], is analogous to the "Central Limit Theorem" for sample averages where they become Normally distributed as sample size increases no matter what distribution the individual values possess.

A superposition example is in **Section 4** where think times are drawn from a **Uniform** distribution as depicted in **Figure 8**. This figure illustrates that as the number of load tool threads, N, drawing think times independently from their individual **Uniform** distributions in **Figure 8a** on the left increase, the collection of them yields time intervals between requests that are **Negative-Exponentially** distributed as depicted in **Figure 8b** on the right.

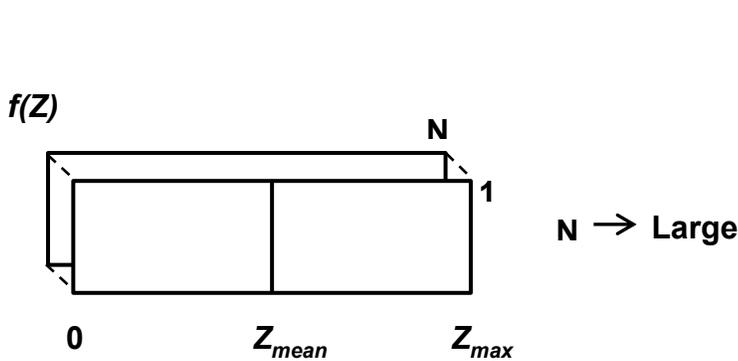 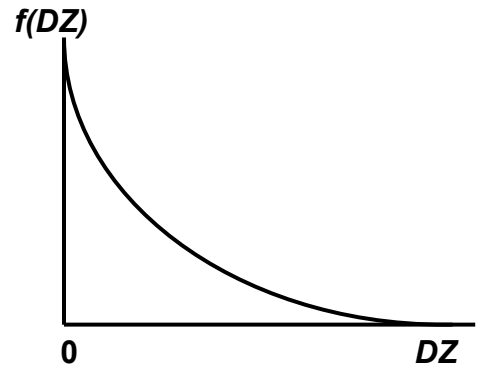

Figure 8a: N Uniformly Distributed Z Values          Figure 8b: Neg-Exp Time Intervals Between Requests

The **Uniform** distribution is used in the **Section 4** example for two reasons; it has a defined range of values to draw from and **JMeter**, the load tool selected, does not have a **Negative-Exponential** distribution option. **JMeter** likely lacks this option because that distribution is unbounded and, in **JMeter**, if a user thread draws a think time that is scheduled to expire after the test finishes, that thread remains alive for the fully scheduled time-period instead of being terminated when the test run timer expires.

### 3.3 Closed Loop Feedback

Another property of the load tool's user process threads that can cause request launch distortion is feedback resulting from the permanent closed loops. A lower percentage of think time in the closed loop's round-trip time (**RT = Z + R**) lessens the think time probability distribution's ability to maintain the desired asynchronous user thread pool with time interval between requests **CoV** ~ 1.0. This is because response times are primarily composed of time in the **SUT** which varies significantly across the set of web events being tested. Response times for simple web events tend to be small while those for complex queries are typically large. The example load test in **Section 4** reflects this diversity with response time coefficient of variation (**CoV** for **R** not **D**) values for all tests performed (**CoV$_R$**) significantly greater than 1.0.

The pie chart on the left side of **Figure 9**, **Figure 9a**, contains percentages of **Z** and **R** where **Z** drives the launch time pattern for the example load test in **Section 4**. The pie chart on the right side, **Figure 9b**, provides an indication of when web request launch timing is impacted by the response time portion of **RT** for the same example.

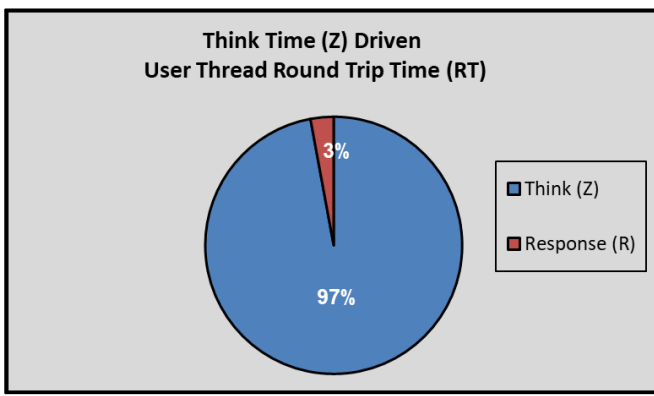 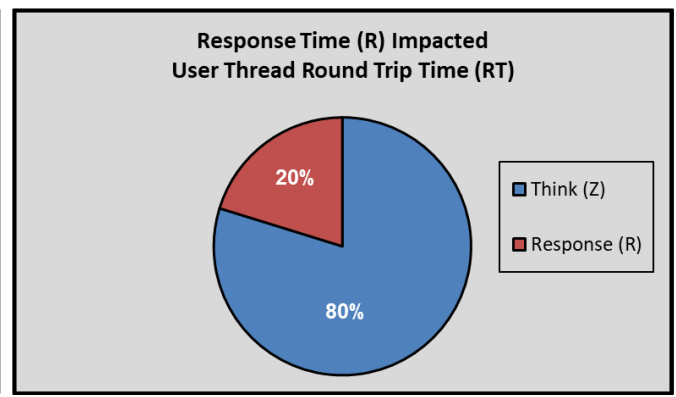

Figure 9a: Think Time Driven RT          Figure 9b: Response Time Impacting RT Timing

This response time impact can happen with real users too, but they are not hard wired and come in and out of the active user pool. As a population, real users can join the active user pool and overload the target system, causing it to crash. The load generator's process thread pool is self-throttling and will eventually stop making requests when all threads are waiting for responses.

One way to alleviate this problem is to focus on traffic mix instead of virtual user count and increase the load tool thread count enough to make the tread pool think time driven. Traffic volume is adjusted using think times that maintain the correct transaction mix with number of active users supported calculated as described in **Section 4.3**.

**4.0 Example Load Test**

The load test included with the **web-generator-toolkit** is used as the example for this paper. The toolkit documentation shows the practitioner how to set the load generator's data collection options, what test run output reports contain, and which report provides the time interval between requests **CoV** information. The testing setup is intended to answer the question; "How many user threads does it take for the time intervals between requests to become statistically independent from each other, $CoV_{DRT}$ ~ 1.0, when each thread is drawing its think times from a **Uniform** distribution?" As will be shown, the example also contains the situation where the $CoV_{DRT}$ diverges, $CoV_{DRT}$ > 1.0, as the user thread count increases.

**4.1 Load Test Description and Results**

The application being tested is a web site used by citizens to obtain state government statistics which is being reconfigured from standalone servers to a virtualized load sharing environment. The load tool is **JMeter** and the **web-generator-toolkit** is used to perform the analysis. **Figure 10** contains the list of web events being tested specified by name and purpose.

| GOV Web Site – Web Page GET / POST Events | |
|---|---|
| **Web Page Name** | **Purpose** |
| 010_Home | Home Page |
| 012_Home_jpg | Background Image |
| 020_Dept | Department Information |
| 022_Dept_jpg | Department Image |
| 030_Demographics | Demographic Information |
| 040_Statistics | Summary Statistics |

Figure 10: GOV Web Site Events Load Tested

**Figure 11a** is a topological view of the load testing environment showing load generator, network interfaces, F5 Load Balancer, and the virtualized Blade Server arrangement with GOV virtual servers "1" and "2" as the **SUT**.

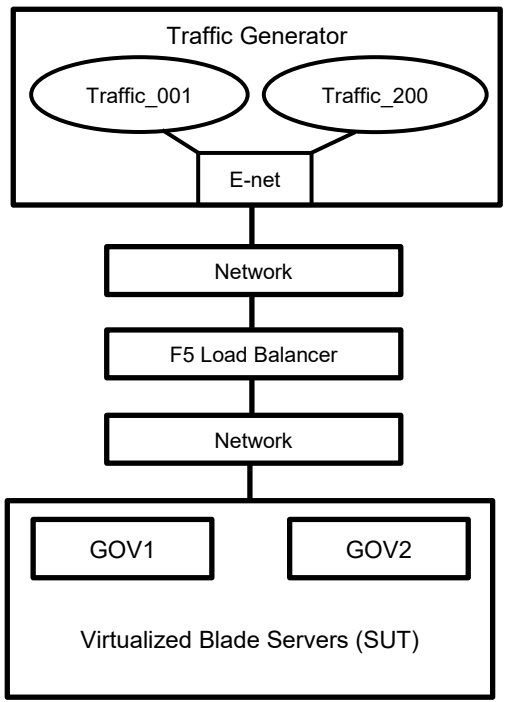
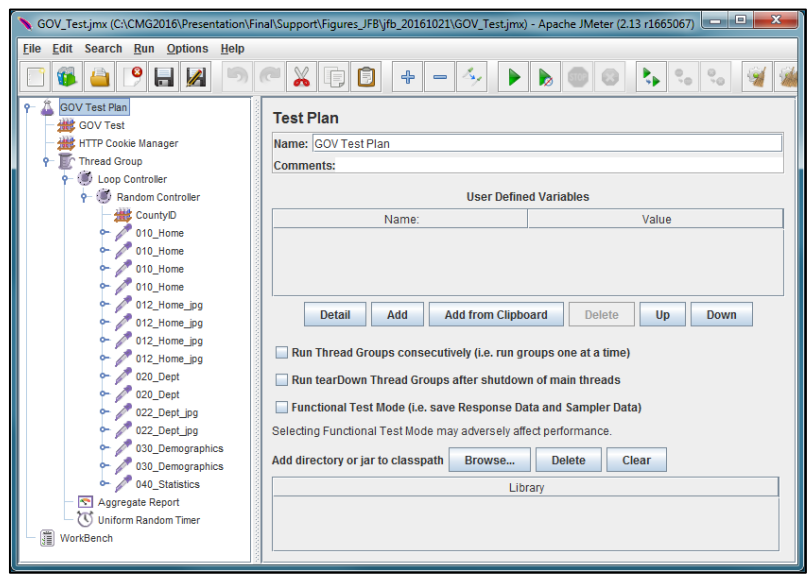

Figure 11a: Load Testing Configuration   Figure 11b: JMeter Load Testing Script

**Figure 11b** displays the **JMeter** load tool script being used to perform the tests. It indicates that web pages are selected in random order, multiple instances of some web events are used to adjust the traffic mix, and one **Uniform** Random Timer performs think time generation. Details regarding test objectives and performance results are contained in [**BRAD12**] and [**BRAD16a**] as well as the **web-generator-toolkit** documentation.

The **CoV<sub>DRT</sub>** convergence to 1 testing setup and methodology is:
1. Two hundred (200) **JMeter** threads are used for all tests,
2. Load is increased from test to test by reducing the **mean** think time,
3. Seven 25-minute tests are performed:
    a. The first 2 and last 3 minutes of each test are excluded from the analysis.
    b. Lowest to highest traffic load test numbers – 1800, 1830, 1900, 1930, 2000, 2030, 2100.

**Figure 12** is a summary of results table containing user thread count (**N**), transaction rate (**Tps**), think time (**Z**) and response time (**R**) **mean** and **standard deviation** as well as associated **CoV<sub>R</sub>** from a load generator perspective for all seven tests performed. **Tps** rates range from 15.91 Trans/Sec for test run 1800 to a maximum of 159.16 Trans/Sec for test run 2100 and **CoV<sub>R</sub>** values range from 2.11 to 3.08.

| | JMeter Load Generator Statistics | | | | | |
|---|---|---|---|---|---|---|
| | **Threads** | **Tps** | **Milliseconds** | | | **CoV$_R$** |
| **Test Run** | **N** | **Trans / Sec** | **Z$_{mean}$** | **R$_{mean}$** | **R$_{sdev}$** | **R$_{sdev}$ / R$_{mean}$** |
| 1800 | 200 | 15.91 | 12500 | 53 | 155 | 2.89 |
| 1830 | 200 | 31.73 | 6250 | 53 | 156 | 2.95 |
| 1900 | 200 | 46.79 | 4200 | 54 | 167 | 3.08 |
| 1930 | 200 | 60.26 | 3250 | 59 | 179 | 3.03 |
| 2000 | 200 | 77.56 | 2500 | 75 | 229 | 3.06 |
| 2030 | 200 | 118.03 | 1563 | 134 | 357 | 2.67 |
| 2100 | 200 | 159.16 | 1000 | 254 | 536 | 2.11 |

Figure 12: GOV Test – Traffic and Response Time Summary

**Figure 12** only contains information about the offered load and response time but what about the launch time pattern of the web requests that is being discussed? The **web-generator-toolkit** provides that information in its Inter-arrival time statistics report. **Figure 13** is the test run 1830 version of that report which contains the time between requests coefficient of variation, **CoV<sub>DRT</sub>** , in the "cv" column. Note that not only is the "cv" for Total = 1 but the "cv" for each web event is also close to 1.

| Inter-arrival Summary Statistics (ms) - select_1830_AggRpt_120_1200 Thursday 03/15/2012 | | | | | | | | | | |
|---|---|---|---|---|---|---|---|---|---|---|
| label | n | tps | median | mean | sdev | cv | p90 | p95 | p99 | min | max |
| 010_Home | 10026 | 8.36 | 82 | 119.68 | 119.18 | 1 | 278 | 358 | 555 | 0 | 1201 |
| 012_Home_jpg | 10205 | 8.51 | 81 | 117.57 | 116.97 | 0.99 | 270 | 352 | 527 | 0 | 1145 |
| 020_Dept | 4975 | 4.15 | 172 | 241.15 | 232.47 | 0.96 | 555 | 710 | 1038 | 0 | 2285 |
| 022_Dept_jpg | 5068 | 4.22 | 166 | 236.76 | 235.35 | 0.99 | 546 | 712 | 1080 | 0 | 2143 |
| 030_Demographics | 5220 | 4.36 | 161 | 229.59 | 229.04 | 1 | 519 | 662 | 1071 | 0 | 2208 |
| 040_Statistics | 2573 | 2.15 | 332 | 465.92 | 458.19 | 0.98 | 1062 | 1419 | 2054 | 0 | 3660 |
| Total | 38072 | 31.73 | 22 | 31.52 | 31.59 | 1 | 73 | 94 | 143 | 0 | 435 |

Figure 13: Inter-arrival time statistics for test 1830.

These "cv" results were obtained by applying the technique described in the "Real Web User Request Timing" box contained in **Section 2.3** to the test run 1830 **JMeter** Aggregate Report output file. **Figure 14** contains the first few records of this output file. The left column, "TimeStamp (ms)", is the query launch time of the event in **Unix time** expressed in milliseconds. The time between requests "Inter-arrival" times are produced by sorting the file in timestamp order and calculating the differences between adjacent timestamp values. For example, the difference between the first two events listed in **Figure 14** is 29 milliseconds (1331861523145 – 1331861523116 = 29). This calculation is repeated at the web event name level, e.g., 010_Home. The **mean** and **standard deviation** of these differences are computed along with their coefficient of variation and reported as shown in **Figure 13**.

| TimeStamp (ms) | R (ms) | Web Event Name | Response Code | Response Message | User Thread | Data Type | Success | Byte Count | R (1st Byte) (ms) |
|---|---|---|---|---|---|---|---|---|---|
| 1331861523116 | 9 | 010_Home | 200 | OK | Thread Group 1-97 | text | TRUE | 17991 | 7 |
| 1331861523145 | 9 | 010_Home | 200 | OK | Thread Group 1-127 | text | TRUE | 17991 | 7 |
| 1331861523160 | 5 | 022_Department_jpg | 200 | OK | Thread Group 1-198 | bin | TRUE | 31541 | 3 |
| 1331861523166 | 9 | 020_Department | 200 | OK | Thread Group 1-8 | text | TRUE | 26632 | 6 |
| 1331861523167 | 25 | 012_Home_jpg | 200 | OK | Thread Group 1-179 | bin | TRUE | 141907 | 2 |
| 1331861523169 | 26 | 012_Home_jpg | 200 | OK | Thread Group 1-87 | bin | TRUE | 141907 | 7 |
| 1331861523213 | 5 | 022_Department_jpg | 200 | OK | Thread Group 1-110 | bin | TRUE | 31541 | 3 |
| 1331861523283 | 14 | 012_Home_jpg | 200 | OK | Thread Group 1-80 | bin | TRUE | 141907 | 3 |
| 1331861523306 | 15 | 012_Home_jpg | 200 | OK | Thread Group 1-52 | bin | TRUE | 141907 | 3 |
| 1331861523330 | 5 | 022_Department_jpg | 200 | OK | Thread Group 1-95 | bin | TRUE | 31541 | 3 |
| 1331861523355 | 7 | 020_Department | 200 | OK | Thread Group 1-168 | text | TRUE | 26632 | 5 |
| 1331861523466 | 15 | 012_Home_jpg | 200 | OK | Thread Group 1-29 | bin | TRUE | 141907 | 3 |
| 1331861523595 | 10 | 010_Home | 200 | OK | Thread Group 1-31 | text | TRUE | 17991 | 8 |
| 1331861523620 | 637 | 040_Statistics | 200 | OK | Thread Group 1-72 | text | TRUE | 110193 | 610 |

**Figure 14: JMeter Aggregate Report Listener Output File – 1830_AggRpt_120_1200.csv**

The **web-generator-toolkit** documentation describes how the **JMeter** Aggregate Report Listener is configured to produce the output contained in **Figure 14**.

**4.2 The Number of User Threads Required for CoV$_{DRT}$ ~ 1.0**

The use of a fixed user thread count (**N** = 200) and a systematic decrease in the **mean** think time (**Z**) to drive up the load may appear to be an unusual approach but it helps answer the question posed at the beginning of this section. How many user threads does it take for the time intervals between requests to become statistically independent from each other when each thread is drawing its think times from a **Uniform** distribution?

Since all seven tests are run with two hundred user threads (**N** = 200) the time interval between requests for any grouping of specific thread sets can be analyzed. The important fields in **Figure 14** for this discussion are the first column, "TimeStamp (ms)", and the sixth column, "User Thread". The "TimeStamp (ms)" values are the web request launch times and the "User Thread" labels indicate which thread is making the web request. For example, the "TimeStamp (ms)" value for the first row of data is "1331861523116" and its corresponding "User Thread" is "Thread Group 1-97".

Given the record layout, this author wrote a Perl script which extracts all the records of the desired "User Thread" grouping. The **web-generator-toolkit** analysis script is run against each grouping where it sorts the records in "TimeStamp (ms)" order, calculates the differences in time between adjacent records, computes the **mean** and **standard deviation** of those differences, and produces the **standard deviation** to **mean** ratio, i.e., the **CoV$_{DRT}$**. The analysis is performed for one thread using "Thread Group 1-1" records, two threads by combining "Thread Group 1-1" and "Thread Group 1-2" records. The process is repeated for thread counts three through two hundred by adding one additional thread each time. **Figure 15** and **Figure 16** contain a selected set of these thread count analyses for the specific test runs indicated, e.g., 1830.

**Figure 15** summarizes the results of this procedure for test run 1830. The table on the right of this figure contains thread count, transaction count and rate, the **mean** and **standard deviation** of the time intervals (milliseconds) between requests, and their ratio, i.e., **CoV$_{DRT}$**. The graph on the left is a plot of the **CoV$_{DRT}$** as a function of thread group size. The table on the right indicates the single thread yields a **CoV$_{DRT}$** = .56 which is very close to the $1/\sqrt{3}$ of a **Uniform** distribution with its location parameter set to zero. The graph indicates an independent request pattern starts at around 10 threads, settles in at approximately 50 of them and stays that way through the maximum of 200.

$$Coefficient\ Of\ Variation\ (CoV_{DRT})$$
$$Test\ Run\ 1830$$

$$CoV_{DRT} = \frac{DRT_{sdev}}{DRT_{mean}}$$

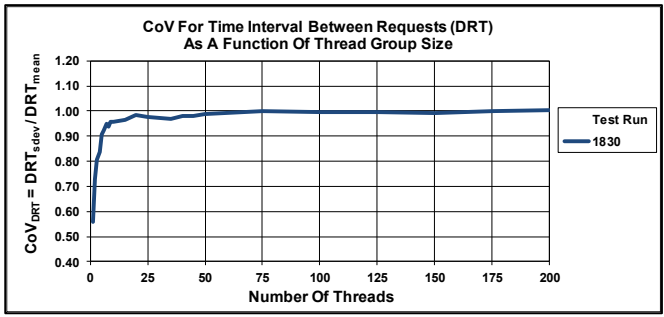

| Test Run | 1830 | | | Time Interval Between Requests (ms) | | |
|---|---|---|---|---|---|---|
| Threads | Trans | Tps | DRT$_{mean}$ | DRT$_{sdev}$ | CoV$_{DRT}$ | |
| 1 | 187 | 0.16 | 6391.96 | 3555.83 | 0.56 | |
| 2 | 360 | 0.30 | 3323.87 | 2423.37 | 0.73 | |
| 3 | 550 | 0.46 | 2175.62 | 1751.27 | 0.80 | |
| 4 | 747 | 0.62 | 1603.27 | 1339.43 | 0.84 | |
| 5 | 939 | 0.78 | 1275.45 | 1157.44 | 0.91 | |
| 6 | 1122 | 0.94 | 1067.42 | 986.39 | 0.92 | |
| 7 | 1317 | 1.10 | 909.37 | 862.32 | 0.95 | |
| 8 | 1509 | 1.26 | 794.09 | 745.60 | 0.94 | |
| 9 | 1694 | 1.41 | 707.37 | 676.15 | 0.96 | |
| 10 | 1875 | 1.56 | 639.65 | 611.60 | 0.96 | |
| 15 | 2877 | 2.40 | 416.95 | 402.70 | 0.97 | |
| 20 | 3846 | 3.21 | 311.90 | 306.66 | 0.98 | |
| 25 | 4814 | 4.01 | 249.23 | 243.05 | 0.98 | |
| 30 | 5776 | 4.81 | 207.72 | 202.35 | 0.97 | |
| 35 | 6710 | 5.59 | 178.80 | 173.23 | 0.97 | |
| 40 | 7685 | 6.41 | 156.12 | 152.82 | 0.98 | |
| 45 | 8606 | 7.17 | 139.41 | 136.55 | 0.98 | |
| 50 | 9566 | 7.97 | 125.42 | 124.02 | 0.99 | |
| 75 | 14292 | 11.91 | 83.95 | 83.88 | 1.00 | |
| 100 | 19089 | 15.91 | 62.86 | 62.55 | 1.00 | |
| 125 | 23864 | 19.89 | 50.28 | 50.03 | 1.00 | |
| 150 | 28580 | 23.82 | 41.99 | 41.60 | 0.99 | |
| 175 | 33276 | 27.73 | 36.06 | 36.00 | 1.00 | |
| 200 | 38072 | 31.73 | 31.52 | 31.59 | 1.00 | |

**Figure 15:** Time Interval Between Requests CoV$_{DRT}$ Vs Thread Group Size for Test Run 1830

The **Figure 16** table and graph contain a summary of the **CoV$_{DRT}$** values for all seven tests. They both show that all test runs have a single thread **CoV$_{DRT}$** very close to $1/\sqrt{3}$ and, except for the last two tests, a **CoV$_{DRT}$** ~ 1.0 beginning around 10 threads, settling in at about 50 threads, and continuing through the full group of 200. Test Run 2030 and Test Run 2100, however, diverge when their thread group is greater than 10 in size with Test Run 2100 reaching a **CoV$_{DRT}$** = 1.17 for 200 threads.

$$Coefficient\ Of\ Variation\ (CoV_{DRT})$$
$$For\ All\ Seven\ Test\ Runs$$

$$CoV_{DRT} = \frac{DRT_{sdev}}{DRT_{mean}}$$

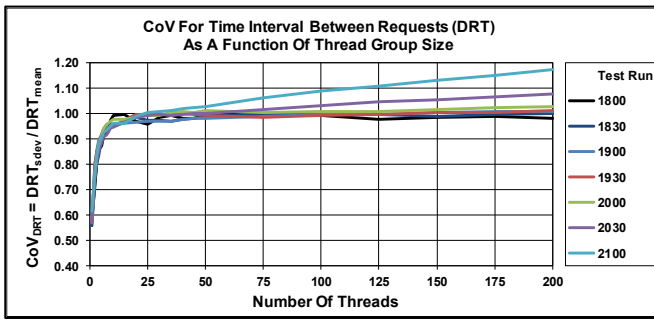

| Test Run | 1800 | 1830 | 1900 | 1930 | 2000 | 2030 | 2100 |
|---|---|---|---|---|---|---|---|
| Threads | CoV$_{DRT}$ = DRT$_{sdev}$ / DRT$_{mean}$ | | | | | | |
| 1 | 0.59 | 0.56 | 0.57 | 0.57 | 0.59 | 0.57 | 0.61 |
| 2 | 0.71 | 0.73 | 0.79 | 0.80 | 0.76 | 0.75 | 0.76 |
| 3 | 0.81 | 0.80 | 0.85 | 0.85 | 0.83 | 0.81 | 0.83 |
| 4 | 0.85 | 0.84 | 0.90 | 0.89 | 0.88 | 0.85 | 0.88 |
| 5 | 0.87 | 0.91 | 0.91 | 0.92 | 0.91 | 0.88 | 0.91 |
| 6 | 0.91 | 0.92 | 0.94 | 0.94 | 0.94 | 0.91 | 0.92 |
| 7 | 0.92 | 0.95 | 0.94 | 0.93 | 0.96 | 0.92 | 0.93 |
| 8 | 0.94 | 0.94 | 0.95 | 0.94 | 0.96 | 0.93 | 0.94 |
| 9 | 0.97 | 0.96 | 0.95 | 0.95 | 0.97 | 0.94 | 0.95 |
| 10 | 0.99 | 0.96 | 0.95 | 0.95 | 0.98 | 0.95 | 0.96 |
| 15 | 1.00 | 0.97 | 0.96 | 0.97 | 0.98 | 0.97 | 0.96 |
| 20 | 0.97 | 0.98 | 0.97 | 1.00 | 0.99 | 0.98 | 0.99 |
| 25 | 0.96 | 0.98 | 0.97 | 1.00 | 0.99 | 0.99 | 1.01 |
| 30 | 0.98 | 0.97 | 0.97 | 1.00 | 0.99 | 1.00 | 1.01 |
| 35 | 0.99 | 0.97 | 0.97 | 0.99 | 1.01 | 0.99 | 1.01 |
| 40 | 0.98 | 0.98 | 0.98 | 1.00 | 1.01 | 1.00 | 1.02 |
| 45 | 0.98 | 0.98 | 0.98 | 0.99 | 1.01 | 1.00 | 1.02 |
| 50 | 0.99 | 0.99 | 0.98 | 0.99 | 1.01 | 1.00 | 1.03 |
| 75 | 1.00 | 1.00 | 0.99 | 0.99 | 1.01 | 1.02 | 1.06 |
| 100 | 0.99 | 1.00 | 1.00 | 0.99 | 1.01 | 1.03 | 1.09 |
| 125 | 0.98 | 1.00 | 1.00 | 1.00 | 1.01 | 1.05 | 1.11 |
| 150 | 0.98 | 0.99 | 1.00 | 1.00 | 1.02 | 1.05 | 1.13 |
| 175 | 0.99 | 1.00 | 1.01 | 1.01 | 1.02 | 1.07 | 1.15 |
| 200 | 0.98 | 1.00 | 1.01 | 1.01 | 1.03 | 1.08 | 1.17 |

**Figure 16:** Time Interval Between Requests CoV$_{DRT}$ Vs Thread Group Size for All Seven Test Runs

What caused the divergence from a **CoV$_{DRT}$** ~ 1.0 as thread group size increased beyond 10 for these two tests? The answer is associated with the closed loop feedback discussed in **Section 3.3**.

### 4.3 Closed Loop Feedback Issues and Nuances

The response time, **R**, portion of round trip time, **RT**, provides a clue why this divergence occurs. **Figure 17** contains a histogram of the **mean** response time to **mean** round trip time ratio, **RoRT**$_{mean}$ = **R**$_{mean}$ / **RT**$_{mean,}$ for all seven test runs. That ratio for the last two test runs, 2030 and 2100, is much larger than for the previous five with a 2100 value of **RoRT**$_{mean}$ = 20.23% (**Figure 9b**). In contrast, test run 2000, the highest traffic level converging to a **CoV**$_{DRT}$ ~ 1.0, has the **RoRT**$_{mean}$ = 2.91% (**Figure 9a**). The high values of **RoRT**$_{mean}$ and **CoV**$_R$ for test run 2030 and 2100 combine to cause their **CoV**$_{DRT}$ numbers to diverge away from 1.0 as thread count increases.

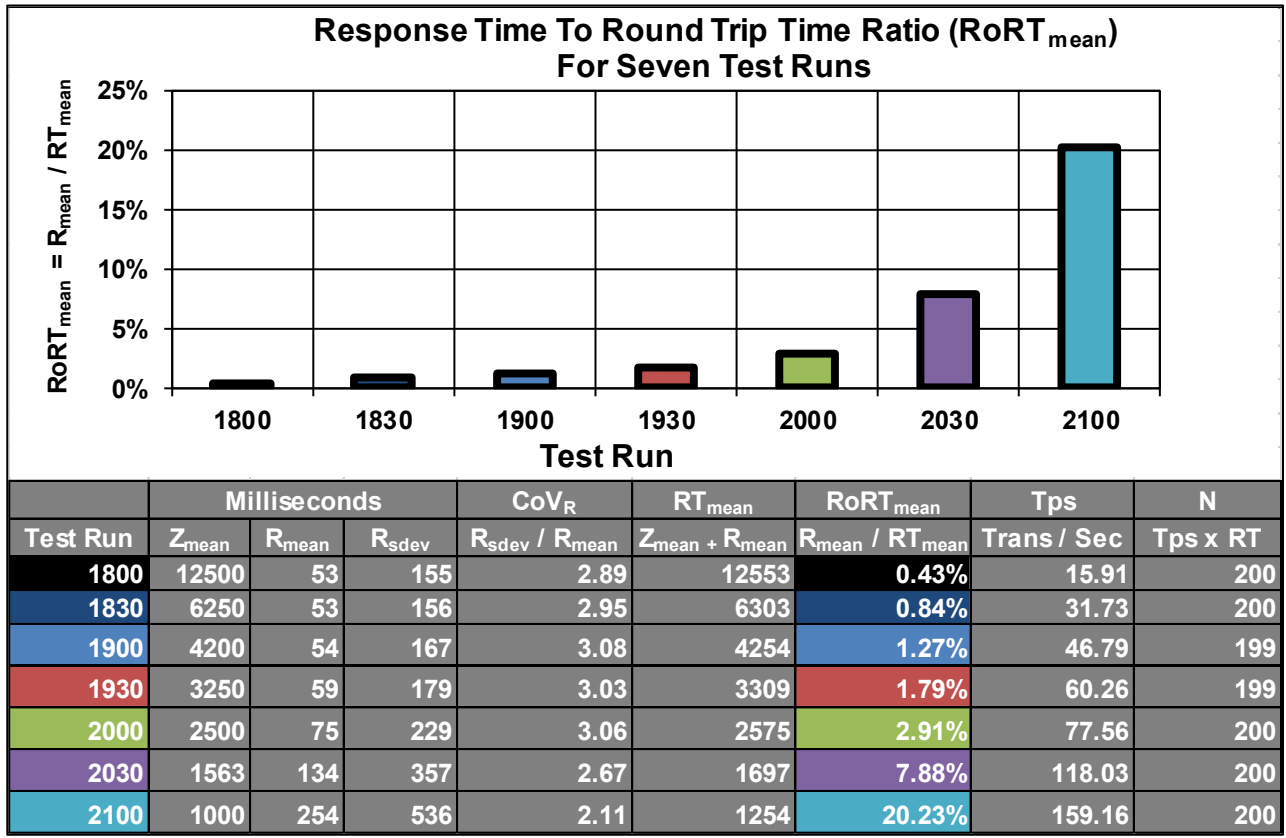

| Test Run | Milliseconds | | | CoV$_R$ | RT$_{mean}$ | RoRT$_{mean}$ | Tps | N |
| | Z$_{mean}$ | R$_{mean}$ | R$_{sdev}$ | R$_{sdev}$ / R$_{mean}$ | Z$_{mean}$ + R$_{mean}$ | R$_{mean}$ / RT$_{mean}$ | Trans / Sec | Tps x RT |
|---|---|---|---|---|---|---|---|---|
| 1800 | 12500 | 53 | 155 | 2.89 | 12553 | 0.43% | 15.91 | 200 |
| 1830 | 6250 | 53 | 156 | 2.95 | 6303 | 0.84% | 31.73 | 200 |
| 1900 | 4200 | 54 | 167 | 3.08 | 4254 | 1.27% | 46.79 | 199 |
| 1930 | 3250 | 59 | 179 | 3.03 | 3309 | 1.79% | 60.26 | 199 |
| 2000 | 2500 | 75 | 229 | 3.06 | 2575 | 2.91% | 77.56 | 200 |
| 2030 | 1563 | 134 | 357 | 2.67 | 1697 | 7.88% | 118.03 | 200 |
| 2100 | 1000 | 254 | 536 | 2.11 | 1254 | 20.23% | 159.16 | 200 |

Figure 17: Response Time to Round Trip Time Ratio

This situation highlights the importance of using a large enough thread group to maintain a small **RoRT**$_{mean}$ ratio, when the **CoV**$_R$ is significantly greater than 1.0.

The last column of **Figure 17** is interesting because it shows the number of user threads can be computed by multiplying the transaction rate by the **mean** round trip time (**N** = **Tps** x **RT**$_{mean}$). This calculation can also be used to estimate number of real active users supported if the user's web page rate is substituted for the load generator's GET/POST transaction rate.

### 5.0 Summary

A load generator's fixed user thread pool can sync up and produce spiky request traffic which results in a misrepresentation of workload flow and erroneous application response time statistics. There are many ways the load generator's process thread pool can distort the request pattern. Three common ways for this to happen that the discussion in **Section 3** details include, heavily loading the traffic generator, using think time algorithms which cause threads to cluster behind the long response time events, and creating closed loop feedback with the **SUT**. When present, any of these distortions can cause the load generator to launch queries in bunches rather than independently from each other like a population of real users.

The independent nature of real user requests was conceptually formalized many years ago and those concepts can be demonstrated with an illustration like the one provided in **Section 2**. That sampling exercise shows that calculating the $CoV_{DRT}$ is the most strait forward way to measure how well the load generator is representing the independence of real user behavior to the system under test, **SUT**. This simple calculation, based on differences between request launch times is all the is required.

The example load test in **Section 4**, taken from the **web-generator-toolkit**, illustrates how the $CoV_{DRT}$ calculation is performed and used to evaluate when good web request timing is being achieved. This example illustrates both convergence to the desired $CoV_{DRT}$ ~ 1 as well divergence away from that value.

The divergence case raises an important question. How robust is the pure virtual user model so many load testing professionals rely on when request independence is such an important factor in the real world? Unfortunately, few practitioners are aware of the $CoV_{DRT}$ and how significant it is for determining the quality of traffic produced by the load generator. Reporting the $CoV_{DRT}$ as a standard metric in load test results adds significant credibility to the effort and if the $CoV_{DRT}$ ~ 1.0 the results are much more likely to match live application performance.

## Copyrights and Trademarks

## Glossary

**CoV:** Coefficient of variation (Standard Deviation / Mean)

**$CoV_{DRT}$** : CoV of time interval between requests based on round trip time ($DRT_{sdev}$ / $DRT_{mean}$)

**$CoV_R$** : CoV of response time ($R_{sdev}$ / $R_{mean}$)

**D:** Time interval between requests

**DRT:** Time interval between requests based on round trip time (Z + R)

**DZ:** Time interval between requests based on think time (Z)

**Mean**: (average): $= \frac{\sum_{i=1}^{n} x_i}{n}$, where $x_i = i^{th}$ sample value and $n =$ sample size.

**N:** Number of load tool process (user) threads

**R:** Response time

**$RoRT_{mean}$:** Mean response time divided by mean round trip time ($R_{mean}$ / $RT_{mean}$)

**RT:** Round trip time (Z + R)

**Standard Deviation (sdev):** $= \sqrt{\frac{\sum_{i=1}^{n}(x_i - \bar{x})^2}{n-1}}$ where $\bar{x} = \frac{\sum_{i=1}^{n} x_i}{n}$, $x_i = i^{th}$ sample value, and $n =$ sample size.

**SUT:** System Under Test

**Tps:** Transactions per second

**Z:** Think time